\documentstyle[11pt,iau185,twoside,epsf]{article}

\begin{document}

\title{Asteroseismology and forced oscillations of \object{HD 209295}, the
first member of two classes of pulsating star}

\author{G. Handler,\altaffilmark{1} L. A. Balona,\altaffilmark{1}
R. R. Shobbrook,\altaffilmark{2,3} C. Koen,\altaffilmark{1}
A. Bruch,\altaffilmark{4} \\ E. Romero-Colmenero,\altaffilmark{1}
A. A. Pamyatnykh,\altaffilmark{5} B. Willems,\altaffilmark{6}
L. Eyer,\altaffilmark{7,8}\\ D. J. James,\altaffilmark{9,10,11,12}
T. Maas,\altaffilmark{7} L. A. Crause\altaffilmark{13}}

\altaffiltext{1}{South African Astronomical Observatory, P.O. Box 9,
Observatory 7935, South Africa}
\altaffiltext{2}{P. O. Box 518, Coonabarabran, N.S.W 2357, Australia}
\altaffiltext{3}{Research School of Astronomy and Astrophysics,
Australian National University, Weston Creek P.O., ACT 2611, Australia}
\altaffiltext{4}{Laboratorio Nacional de Astrofisica, Itajuba, Brazil}
\altaffiltext{5}{Copernicus Astronomical Center, ul. Bartycka 18, 00-716 Warsaw,
Poland}
\altaffiltext{6}{Department of Physics and Astronomy, The Open University,
Walton Hall, Milton Keynes MK7 6AA, UK}
\altaffiltext{7}{Instituut voor Sterrenkunde, Katholieke Universiteit
Leuven, B-3001 Heverlee, Belgium}
\altaffiltext{8}{Astrophysical Sciences Dept., Princeton University,
Princeton, New Jersey 08544, USA}
\altaffiltext{9}{Observatoire de Gen\`{e}ve, Chemin des Maillettes 51,
CH-1290 Sauverny, Switzerland}
\altaffiltext{10}{Laboratoire d'Astrophysique, Observatoire de Grenoble,
Universit\'{e}
Joseph Fourier, B.P. 53, F-38041, Grenoble Cedex 9, France}
\altaffiltext{11}{School of Physics \& Astronomy, University of St
Andrews, North Haugh, St Andrews, FIFE, KY16 9SS, United Kingdom}
\altaffiltext{12}{475 N. Charter Street, 5534 Sterling Hall, Madison, WI
53706-1582, USA}
\altaffiltext{13}{Department of Astronomy, University of Cape Town, Rondebosch 7700,
South Africa}

\keywords{Stars: oscillations, Stars: variables: $\delta$ Sct, 
Stars: individual: \object{HD 209295}, Stars: binaries: close, Stars: binaries:
spectroscopic -- Stars: neutron}

\section*{Abstract}

We report the discovery of both intermediate-order gravity mode and
low-order pressure mode pulsation in the same star, \object{HD 209295}. It
is therefore both a $\gamma$ Doradus and a $\delta$ Scuti star, which
makes it the first confirmed member of two classes of pulsating star.

This object is located in a close binary system with an unknown, but
likely degenerate companion in an eccentric orbit, and some of the
$\gamma$ Doradus pulsation frequencies are exact integer multiples of the
orbital frequency. We suggest that these pulsations are tidally excited.
\object{HD 209295} may be the progenitor of an intermediate-mass X-Ray binary.

\section{Introduction}

The variability of \object{HD 209295} was discovered by the HIPPARCOS
mission (ESA 1997). Handler (1999) performed a frequency analysis of these
data and found multiperiodic variability only explicable by $\gamma$
Doradus-type pulsation. As this object is also located in the $\delta$
Scuti instability strip, Handler \& Shobbrook (2002) examined it for
short-term variability and indeed discovered short-period pulsations
superposed on the $\gamma$ Doradus light curve.

It was then clear that \object{HD 209295} was the first confirmed member
of two classes of pulsating star. Consequently, follow-up observations,
including a multi-site campaign were organized, resulting in 280 h of
photometric and 135 h of spectroscopic monitoring. We show an example
light curve in Fig.~1.

\begin{figure}
\plotfiddle{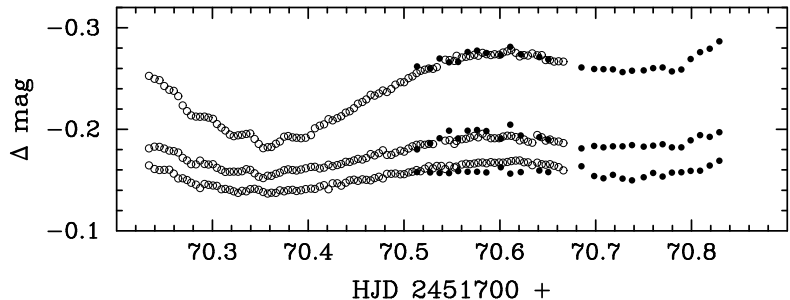}{3.8cm}{0}{90}{90}{-192}{-42}
\caption{An example light curve of the $\gamma$ Doradus/$\delta$ Scuti
star \object{HD 209295}. The open circles are measurements from SAAO and the filled
circles are LNA (Brazil) data. The upper light curves in each panel are
the $V$ data, the middle ones the $(V-I_{\rm c})$ variations and the
lowest are the $(B-V)$ light curves. Note the multiperiodic slow and rapid
light variability and the corresponding colour changes.}
\end{figure}

\section{Results}

The analysis of our measurements yielded several surprises. Firstly, it
turned out that the star shows quite large radial velocity variations, in
excess of 100 km/s. This was reconciled with binary motion: \object{HD
209295} is a single-lined spectroscopic binary with an orbital period of
3.10575~d and a rather high eccentricity of 0.35. The absence of
ellipsoidal variability results in an upper limit of the orbital
inclination of 44$\deg$, which in turns requires a companion mass $>0.98
M_{\sun}$.

We searched for the binary companion which should make itself obvious
through an infrared excess if it was a main-sequence star, but could not
detect it. However, we noticed that TD-1 satellite measurements of
\object{HD 209295} showed quite a large ultraviolet excess, but no binary
model with a hot companion subluminous in the optical can explain the
amount of the excess flux. The high orbital eccentricity and our
constraint on the secondary mass suggest a neutron star as the orbital
companion, but this explanation also has its weaknesses, as the star seems
chemically normal, has normal space motions, and is not an X-Ray source.

Turning back to the pulsations, we noted little correspondence between the
frequency analysis of the photometric and radial velocity measurements.
This is at first sight also surprising, but can be explained by the low
amplitude of the $\delta$ Scuti pulsations and toroidal corrections due to
stellar rotation (Aerts \& Krisciunas 1996) for the $\gamma$ Doradus
modes.

We were able to obtain a mode identification for the two highest-amplitude
photometric modes which are weakly present in the line-profile variations
by using the method by Telting \& Schrijvers (1997). According to this
result, both are $\ell=1, |m|=1$. As we have quite a substantial amount of
time-series colour photometry of \object{HD 209295} at our disposal, we
also attempted mode identifications following Balona \& Evers (1999) and
Koen et al. (1999). However, these results were (at best) inconclusive.

\section{Evidence for forced oscillations}

Another intriguing result from our study was that several of the
photometrically detected variations have frequencies which are exact
integer multiples of the orbital frequency (see Table 1). This raises the
suspicion that these are tidally excited modes.

\begin{table*}
\caption[]{The final multifrequency solution for \object{HD 209295} from
multisite photometry from the year 2000. Pulsational phases for zero
amplitude are given with respect to one time of periastron passage, HJD
2451771.864.}
\begin{center}
\begin{tabular}{cccccc}
\hline
ID & Combination & Frequency & $S/N$ & $V$ Ampl. & $V$ phase\\
 & & (d$^{-1}$) & & (mmag) & ($\deg$)\\
 & & & & $\pm$ 0.2 & \\
\hline
$f_1$ & & 1.12934 $\pm$ 0.00005 & 63.7 & 38.9 & -40.3 $\pm$ 0.3 \\
$f_2$ & & 2.30217 $\pm$ 0.00006 & 49.9 & 28.7 & -116.4 $\pm$ 0.4 \\
$f_3$ & 8$f_{\rm orb}$ & 2.57593 $\pm$ 0.00011 & 23.2 & 13.2 & 67.2 $\pm$ 0.9 \\
$f_4$ & $f_2-f_1$ & 1.17283 $\pm$ 0.00004 & 14.4 & 7.6 & -73.3 $\pm$ 1.5\\
$f_5$ & 9$f_{\rm orb}-f_1$ & 1.76859 $\pm$ 0.00005 & 13.6 & 8.2 & 95.4 $\pm$ 1.4 \\
$f_6$ & 7$f_{\rm orb}$ & 2.25394 $\pm$ 0.00011 & 10.6 & 6.6 & -1.0 $\pm$ 1.8 \\
$f_7$ & 3$f_{\rm orb}$ & 0.96597 $\pm$ 0.00011 & 8.9 & 6.2 & -35.9 $\pm$ 1.9 \\
$f_8$ & 5$f_{\rm orb}$ & 1.60996 $\pm$ 0.00011 & 5.8 & 3.9 & -158.6 $\pm$ 3.0 \\
\vspace{0.5mm}
$f_9$ & 9$f_{\rm orb}$ & 2.89792 $\pm$ 0.00011 & 5.7 & 3.5 & 48.9 $\pm$ 3.3 \\
$f_A$ & & 25.9577 $\pm$ 0.0015 & 5.2 & 1.4 & 162.8 $\pm$ 8.3 \\
\vspace{0.5mm}
$f_B$ & & 13.6873 $\pm$ 0.0018 & 4.2 & 1.2 & -3.4 $\pm$ 9.9\\
\hline
 & $f_{\rm orb}$ & 0.32199 $\pm$ 0.00011 \\
\hline
\end{tabular}
\end{center}
\end{table*}

We note that the photometric colour amplitude ratios of the variations at
orbital harmonic frequencies are consistent with those of pulsations.
Table 1 also shows that these variations are also similarly aligned near
periastron (allowing for a 180\deg\, phase ambiguity which is explained in
this scenario as the dominating components of tidally induced oscillations
should be $\ell=2$, $|m|=2$ modes (e.g. see Kosovichev \& Severnyj 1983).

Consequently, one of us (BW) is currently investigating the case for
tidally excited oscillations of \object{HD 209295} theoretically, with
initially quite encouraging results. We refer to Willems \& Aerts (2001)
and Kumar et al. (1995) for more information on theoretical studies of
tidal excitation of nonradial oscillations.

\section{Conclusions}

We showed that \object{HD 209295} is not only the first member of two classes of
pulsating star (it is both a $\gamma$ Doradus and a $\delta$ Scuti star),
but we have also discovered good evidence for the presence of tidally
excited pulsation modes, whereby the responsible binary companion is
possibly a neutron star.

The confrontation of photometric and spectroscopic mode identification
methods resulted in a reasonable identification for the two dominating
$\gamma$ Doradus modes of the star, whereas the photometric techniques did
not yield meaningful results. The analysis of the suspected tidally
excited modes may give us further insight into this problem.

Despite the many fascinating results obtained from the study of \object{HD 209295},
several open questions still remain. We do not well understand why there
is little correspondence between the photometric and radial velocity
frequency analyses. The nature of the companion of \object{HD 209295} is also
presently a mystery.

Future observational studies of the star which would be worthwhile would
be an even larger multisite photometric and spectroscopic campaign, in
particular with larger telescopes for high-resolution spectroscopy. There
are still further pulsational signals to be found. UV and infrared
spectroscopy would help to find the orbital companion, and an abundance
analysis may reveal an unusual chemical composition caused by the previous
evolution of the star through, e.g. a mass transfer stage.


\begin{references}

\reference Aerts, C., Krisciunas, K., 1996, MNRAS 278, 877

\reference Balona, L. A., Evers, E. A., 1999, MNRAS 302, 349

\reference ESA, 1997, The Hipparcos and Tycho catalogues, ESA SP-1200

\reference Handler, G., 1999, MNRAS 309, L19

\reference Handler, G., Shobbrook, R. R., 2002, MNRAS, in preparation

\reference Koen, C., Van Rooyen, R., Van Wyk, F., Marang, F., 1999, MNRAS,
309, 1051

\reference Kumar, P., Ao, C. O., Quataert, E. J., 1995, ApJ 449, 294

\reference Kosovichev, A. G., Severnyj, A. B., 1983, Pis'ma v 
Astronomicheskii Zhurnal 9, 424

\reference Telting, J. H., Schrijvers, C., 1997, A\&A 317, 723

\reference Willems, B., Aerts, C., 2001, A\&A, in press

\end{references}
\end{document}